\documentclass{WileyMSP-template}

\usepackage{graphicx}
\usepackage{dcolumn}
\usepackage{bm}
\usepackage[utf8]{inputenc}
\usepackage[T1]{fontenc}
\usepackage{hyperref}
\usepackage{amssymb}
\usepackage{amsmath}
\usepackage[scr=rsfso,cal=zapfc,frak=euler,bb=ams]{mathalfa}
\usepackage{caption}
\usepackage{subcaption}
\usepackage{tikz}
\usepackage{xcolor} 
\usepackage{booktabs}
\usepackage{miller}
\usepackage{braket}
\usepackage[nomarkers,figuresonly]{endfloat}

\begin{document}
\pagestyle{fancy}
\rhead{\includegraphics[width=2.5cm]{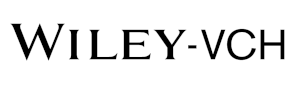}}

\title{Tailoring Nuclear Spins Order With Defects: A Quantum Technology CAD Study}
\maketitle


\author{Gaetano Calogero}
\author{Ioannis Deretzis}
\author{Giuseppe Fisicaro}
\author{Damiano Ricciarelli*}
\author{Rosario Gaetano Viglione}
\author{Antonino La Magna*}

\begin{affiliations}
Consiglio Nazionale delle Ricerche, Istituto per la Microelettronica e Microsistemi (CNR-IMM), Z.I. VIII Strada 5, 95121 Catania, Italy\\
Email Address: damiano.ricciarelli@cnr.it, antonino.lamagna@cnr.it
\end{affiliations}



\begin{abstract}

The full design of relevant systems for quantum applications, ranging from quantum simulation to sensing, is presented using a combination of atomistic methods. A prototypical system features a two-dimensional ordered distribution of spins interacting with out-of-plane spin drivers/probes. It could be realized in wide-bandgap semiconductors through open-volume point defects and functionalized surfaces with low Miller indexes. We study the case of defect electron spins (driver / probe) interacting via hyperfine coupling with $S=1/2$ nuclear spins of H atoms chemisorbed onto \hkl(001) and \hkl(111) 3C-SiC surfaces.  We simulate the system fabrication processes with super lattice kinetic Monte Carlo, demonstrating that epitaxial growth under time-dependent conditions is a viable method for achieving controlled abundance or depletion of near-surface point defects. Quantum features are evaluated by means of extensive numerical analysis at a full quantum mechanical level based on calibrated models of interacting spin systems. This analysis includes both stationary (relative stability of ordered states) and time-dependent (protocols) conditions, achieved varying the model parameters (in our case the atomic structure and the external field). We identify a rich scenario of metastable spin-waves in the quantum simulation setting. The interaction between protocols and variable system configurations could hinder the effectiveness of the preparation/measurement phases.

\end{abstract}


\section{Introduction}
In the last decades, research on quantum systems has entered the realm of application-driven  advancements, while at the same time,  fundamental investigations have become more focused on real materials and structures of interest, and scientific curiosity is enhanced by promises of social impact \cite{Castelvecchi2025, Altman2021, willow2024, aghaee2024interferometric}. So far, quantum supremacy has been particularly demonstrated against difficult, almost impossible, classical computational challenges. However, this is becoming even more apparent today as we can witness a rapid expansion of cases where the "supreme features" of coherently interacting particles and fields offer solutions beyond the classical or semi-classical ones \cite{Daley2022, Hangleiter2023}.
The superior performances of quantum systems in multiple applications and technologies, from sensing to computing, are often supported or predicted by theoretical and computational analysis based on highly idealized models of the corresponding real systems \cite{kitson2024sub,Gu2025}. The related approximations are generally justified by the fundamental assumption that manipulation protocols enable a degree of isolation (in limited space-time scales) of the quantum effects with respect to disturbances due to the surrounding environment \cite{fazio2024PRA}.  

However, in addition to these intrinsic and model-related approximations, which can be properly discussed by appropriate theories and also validated with the aid of experimental studies \cite{seo2016quantum}, we can notice the generic lack of theoretical/computational predictions of the realistic micro-state emerging from the processes used to fabricate the quantum systems. Moreover, the link between potential process simulations and the model applied to describe the desired quantum enhanced performances is usually absent. The missing correlation between process and protocol simulation could be particularly critical for the case of long sequences of control and manipulation procedures necessary to implement complex quantum functionality.  

Thus, we notice that the current stage of the technology computer aided design (TCAD) dedicated to quantum technology (QT) is still early, or at least not sufficiently mature, when compared to tools with analogous scope for traditional technologies. This aspect becomes more important considering that the forthcoming QT breakthrough will be linked to advanced nanoelectronics.  

In this work, we discuss a particular case study of a sufficiently complete Quantum Technology Computer Aided Design (QT-CAD) analysis focused on the cubic polymorph of silicon carbide (3C-SiC). The studied system is composed of electronic spin states, related to open volume defects (namely, $V_{\mathrm{C}} -V_{\mathrm{Si}}$ di-vacancies), which interact through hyperfine coupling with the $S=1/2$ nuclear spins of H atoms chemisorbed onto reconstructed \hkl(001) or \hkl(111) 3C-SiC surfaces. Bulk to near-to-surface defect configurations \cite{zhu_galliananolett2024} are considered to gradually achieve hyperfine coupling intensity, which potentially exceeds locally the nuclear dipole-dipole coupling. 
 
To motivate this choice, we can assert that undoped or unintentionally slightly n-doped silicon carbide is nowadays a material of interest for QTs, with real demonstrators already available as quantum sensors ~\cite{simin2015PhysRevApplied}. However, SiC is mainly a key material for the large-scale production of solid-state devices~\cite{kimoto2014fundamentals} which are integrated into electronic systems with a high social impact (automotive, green energy, low-consumption appliances, etc.). As a positive effect of this huge industrial interest for QT applications, we notice the increasingly efficient technologies that grow and functionalize (e.g. through doping, oxidation, microstructuring, etc.) high-quality SiC substrates or epitaxially regrown thin films with controlled defectiveness. This consideration surely holds for the case of the reference hexagonal 4H-SiC polytype (at the industrial level), but it could also recently be extended to the cubic polytype ~\cite{fisicaro2020genesis,la2021new}. The latter is especially attractive for QTs due to the possibility of being grown by hetero-epitaxy on silicon substrates, thus implementing spin and photon-based quantum functionalities in the same device ~\cite{castelletto2020silicon}.    

\hspace*{0.5cm}The paper is organized as follows: in section 2 we present the effective Hamiltonian for the low-energy manifolds of non-interacting defects and surface states in 3C-SiC materials embedded in external uniform magnetic fields. In section 3 we discuss the atomic simulation results of the formation and manipulation processes aiming at the synthesis of a functionalized material with controlled defects and surface related spin distributions, section 4 is dedicated to the presentation of the rich competing spin phases, which could be stabilized in their collective ground state by the nuclear spin related to the hydrogen coverage of low Miller index surfaces. In section 5 a critical evaluation of the quantum simulator protocols in dependence of the space distribution of the defects is presented, based on full quantum mechanical simulations of time-dependent coupled defect-surface spin models.  
Finally, in section 6 conclusions are drawn, together with a perspective of this study.


\section{\label{sec:baremodels} Spin states of open volume defects and H surface functionalization}

Isolated open volume point-defects (OVPDs, as vacancies, di-vacancies, or anti-site vacancy complexes) in 3C-SiC can generate effective electron non-zero spin states, associated to their lower energy manifold, which in the presence of a constant and uniform external magnetic field $\mathbf{B}$ can be approximately described by the spin Hamiltonian: 
\begin{equation}
\mathcal{H}_{gs} =D S_{z}^2 + E \left( S_{x}^2-S_{y}^2 \right) + \gamma_{e} \mathbf{B} \cdot \mathbf{S} 
\label{eq:ham2V3CSiC}
\end{equation}

\begin{figure}
\begin{center}
\includegraphics[width=0.6\textwidth]{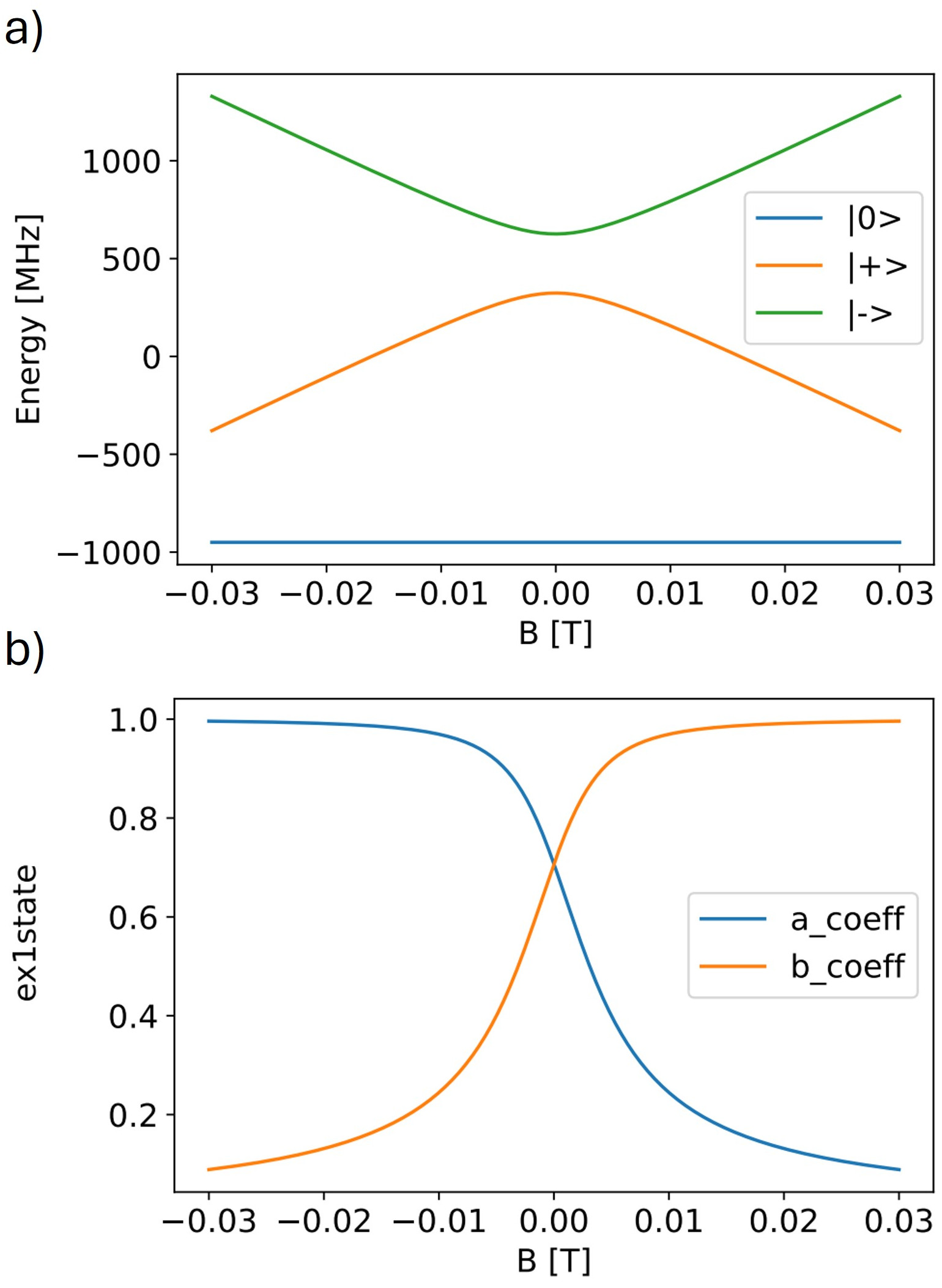}
\end{center}
\caption{(a) Eigenvalues of the model in Eq. \ref{eq:ham2V3CSiC} as a function of the external magnetic field  $B$. (b) Coefficient of the first excited state of the model in Eq. \ref{eq:ham2V3CSiC} as a function of the external magnetic field $B$.}
\label{fig:diagHgs}
\end{figure}

where $\mathbf{S}$ is the S-spin operator, $D$ and $E$ are the components of the Zero-Field Splitting (ZFS) tensor of the defect related ground state and $\gamma_{e}$ is the the gyromagnetic ratio of the electron spin. In principle, for a fixed atomic configuration, the most stable states of OVPD can emerge as charged or neutral and with different associated spin states $\mathbf{S}$ depending on the electrochemical potential (i.e., the sign and the level of doping) of the hosting system \cite{gordon2015defects,fazio2024PRA}. However, experimental realization and analysis can limit this multiplicity to a restricted subset of abundant and interesting cases (such as the $S$ = 1 neutral di-vacancy or the charged $S$ = 3/2 Si vacancy) due to the kinetic and thermodynamic constraints of the processes / substrate combination used to generate OVPDs \cite{son2021jap}.         

\begin{figure*}
    \includegraphics[width=1.0\textwidth]{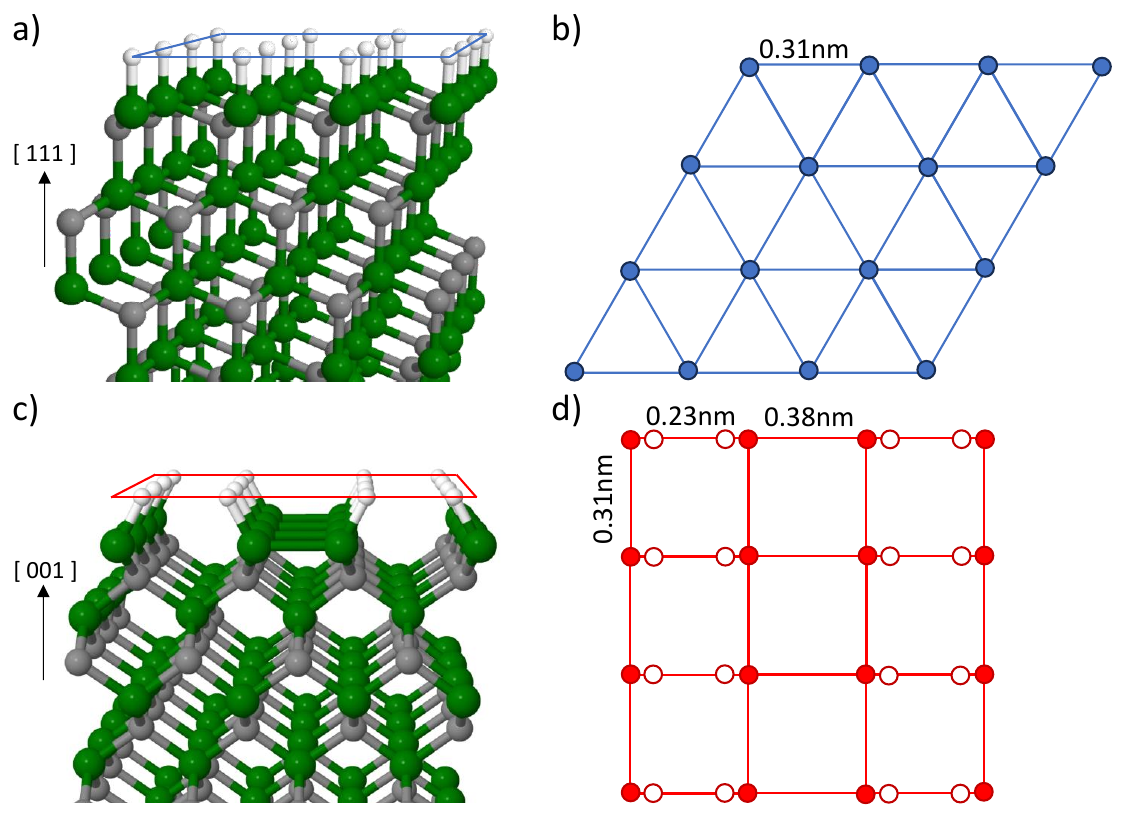}
    \caption{a) Stable H distribution in \hkl(111) surface as obtained by \textit{ab-initio} calculation (white spheres H atoms, gray spheres C atoms, light brown spheres Si atoms). b) H nuclear spins 2D lattice in the SiC \hkl(111) triangular lattice (filled blue circles). c) Stable H distribution in a $2\times 1$ reconstructed \hkl(001) 3C-SiC surface as obtained by \textit{ab-initio} calculation. d) H nuclear spins 2D lattice in the ideal (filled red circles) and $2\times 1$ reconstructed SiC \hkl(001) surfaces (empty red circles). Lattice constants are indicated.}
    \label{fig:H2Dlattices}
\end{figure*}

In the following, with the aid of simulations, we propose an optimized method of process recipe and process parameters for the generation of neutral di-vacancies in 3C-SiC. We focus on these defects because they represent key color centers in silicon carbide, with promising properties for quantum simulators including a high-fidelity character, in analogy to nitrogen-vacancy centers in diamond. As a matter of fact, the neutral 3C-SiC di-vacancies have been frequently detected and fully characterized in epitaxial 3C-SiC and are predicted to be stable over a wide range of growth conditions  \cite{gordon2015defects, christle2017isolated, castelletto2020silicon, Anderson2022}.  The method aims at producing controlled space distributions of variable densities of $V_{\mathrm{C}}-V_{\mathrm{Si}}$ pairs. The bulk neutral di-vacancy in 3C-SiC is experimentally characterized as a $S=1$ spin state with negligible $E \sim 0$ and $D=1328$ MHz \cite{falk2013polytype}, while {\it ab initio} results confirm the close to zero $E$ value (justified also by symmetry considerations based on the stable atomic configuration) but a larger value of $D=1425$ MHz for the ZFS diagonal component \cite{zhu_galliananolett2024}. ZFS parameters of the near-surface defects in Eq. \ref{eq:ham2V3CSiC} have not been estimated yet with magnetic resonance experiments; anyhow, they have been computed with {\it ab initio} methods by Zhu and co-workers \cite{zhu_galliananolett2024}. Their study shows that the asymmetry due to the surface proximity can produce a symmetry breaking in the spin-state with a consequent non-zero $E$ value (in the range of $15-151$ MHz) for defect distances smaller than 1 nm from the reconstructed (100) surface (with a full H coverage) while the $D \sim 1425$ MHz value is similar to the calculated bulk one (less than 1\% variation). Therefore, the excited states for a near surface di-vacancy in 3C-SiC have a mixed character with respect to the eigenstates $\vert \pm 1 \rangle$ of $S_{z}$ which depend on the external magnetic field  (see Figures 1a and 1b)
\begin{equation}
\vert +(B)\rangle= a(B) \vert +1 \rangle + b(B) \vert -1 \rangle
\label{eq:ex1}
\end{equation}
and this character is close to the bulk defect for values of $B$ in the order of few hundreds Gauss (e.g. $a(B)=0.996055$ and $b(B)=0.088737$ at $B=300$ G).      

Exploiting bulk and near-surface defects for quantum functionalities leads to the necessity of a proper surface treatment to obtain sufficiently clean and fully saturated interfaces (i.e. ideally with no residual surface states) with the external environment. Hydrogen-based cleaning and the resulting mono-layer coverage seem an obvious answer to satisfy this requirement, with the additional potential benefit of perfect saturation behavior with respect to other possible 3C-SiC surface terminations (see \cite{zhu_galliananolett2024}). 

However, H nuclear spins ($S=1/2$ for the proton-like most abundant isotope and $S=1$ for the second abundant deuterium stable isotope)  are put into play from the choice of H for the surface treatment. The presence of this 2D interacting distribution of spins could be either a problem or the solution in consideration of the desired quantum application. The Hamiltonian describing the point-like dipolar interaction in the presence of an external magnetic field $B$ is the bare model for, again, the lower energy manifold of this system  

\begin{equation}
\mathcal{H}_{NN} =   \sum_{i} \gamma^n_{i} \mathbf{B} \cdot \mathbf{\sigma}(i) + \frac{\mu_{0}}{4 \pi}\sum_{i,j} \frac{\gamma^n_{i} \gamma^n_{j} }{r_{i,j}^3} \left[\mathbf{\sigma}(i) \cdot  \mathbf{\sigma}(j) -3\left(\mathbf{\sigma}(i) \cdot  \hat{r}_{i,j}\right)  \left( \mathbf{\sigma}(j) \cdot \hat{r}_{i,j}\right) \right] 
\label{eq:dipdip}
\end{equation}

where $\gamma^n_{i}$ is the giromagnetic ratio for the $i_{th}$ nuclear spin $\mathbf{\sigma}(i)$, $\mathbf{r}_{i,j}=r_{i,j} \mathbf{\hat{r}}_{i,j}$ is the vector that connects the position of the $i_{th}$ and $j_{th}$ nuclear spins and $\mu_{0}$ is the vacuum magnetic permittivity.  
The model of Eq. \ref{eq:dipdip} for sufficiently high external magnetic field, i.e. when the rotating wave approximation (RWA) holds,  
\begin{equation}
\gamma^n_{i} B \gg  \frac{\mu_{0}}{4 \pi} \frac{\gamma^n_{i}\gamma^n_{j}}{r_{i,j}^3} 
\end{equation}
is approximated, in a interaction picture with respect to the leading term \cite{cai2013natureQS}, by a $\Delta =1/2$ XXZ model with a position dependent coupling strength

\begin{equation}
\mathcal{H}_{XXZ} =   \frac{\mu_{0}}{4 \pi}\sum_{i,j} \frac{\gamma^n_{i}\gamma^n_{j}}{r_{i,j}^3} \left(1 - 3  (\hat{\mathbf{r}}_{i,j}^B)^2\right) \left[\sigma_z(i) \sigma_z(j) -\Delta \left(\sigma_x(i)\sigma_x(j) + \sigma_y(i)\sigma_y(j) \right)  \right] 
\label{eq:XXZ}
\end{equation}

where $\hat{\mathbf{r}}_{i,j}^B$ is the component of the versor $\hat{\mathbf{r}}_{i,j}$ along the direction of the external magnetic field. Here and in the following we assume that the $X-Y$ plane coincides with the SiC surface where H atoms chemisorb and that the $Z$ axis is aligned along the direction perpendicular to this surface.  

The coupling parameters in the models \ref{eq:dipdip} or \ref{eq:XXZ} depend on the atomic space distribution of the H atoms on the low Miller index surfaces. In Figure \ref{fig:H2Dlattices} examples of H atom arrangements for two of these surfaces (\hkl(001) in Figures \ref{fig:H2Dlattices} a-b) and \hkl(111) in Figures \ref{fig:H2Dlattices} c-d) with the Si-type polarity are shown. We notice that, while in the \hkl(111) case the hydrogen 2D lattice has the same lattice constant of the \hkl(111) 3C-SiC ideal surface, the \hkl(001) stable configurations, calculated with \textit{ab initio} methods (see section S1.1 in Supporting Information), deviate from the ideal square lattice, leading to dimerized hydrogen displacements along one direction.      

\section{\label{sec:process} Processes simulations}
The models \ref{eq:ham2V3CSiC} and \ref{eq:dipdip} or \ref{eq:XXZ} represent the bare Hamiltonians for the key components of our system. The (electronic) defects and nuclear spins interact directly and indirectly (i.e. when interactions are mediated by other spurious spins present in the hosting environment \cite{fazio2024PRA}, \cite{nagy2019high}) by means of hyperfine coupling. Dealing with these interactions is one computational task in our QT-CAD approach. However, the correct predictions obtained through a full model including also interaction Hamiltonians are, in a complete QT-CAD chain, the results of reliable simulations of the atomic microstate emerging from the processes. In this section we discuss an example of such process simulation steps addressing the selective and controlled generation of open volume defects. We also notice that surface functionalization with H atoms can be simulated with these methods \cite{Raciti20231MSSP} \cite{MulSKIPS}. Moreover, more complex hydrogenation kinetics, with the concurrent production of quasi-free-standing graphene coating, can also be modeled \cite{Deretzis2013Nanoscale} with atomic resolution.      


Point-like defect configurations are usually experimentally generated irradiating the sample with energetic (ionic \cite{falk2013polytype,bathenPhysRevB.104.045120}, electronic \cite{son2021jap} or neutrons \cite{akel2023neutrons}) beams. The interaction between the substrate and energetic beams and more specifically the atomic displacements in the resulting collision cascades promote the formation of the whole typology of defects (i.e., vacancy, interstitial and anti-site types), whereas impinging particles and displaced atoms belong to the interstitial class. A more selective creation of defects, unbalanced towards OVPDs, can in principle be obtained directly with growth processes. This fact makes high-purity semi-insulating samples of immediate use for di-vacancy characterization \cite{falk2013polytype,seo2016quantum} but it is barely used for the controlled generation of these defects.  

Recent experiments based on measurements of positron annihilation on epitaxially grown Si substrates have demonstrated that the growth speed is monotonically related to the density of OVPDs in grown films \cite{isa2024jappositron}, with a generated vacancy density of the order of $10^{20}$cm$^{-3}$ for high growth speed ($\sim 5$ nm/s) processes. These concepts can be extended to the 3C-SiC epitaxial growth as we will discuss in the following with the aid of atomistic simulation of the process. 

\begin{figure*}
\begin{center}
    \includegraphics[width=1.0\linewidth]{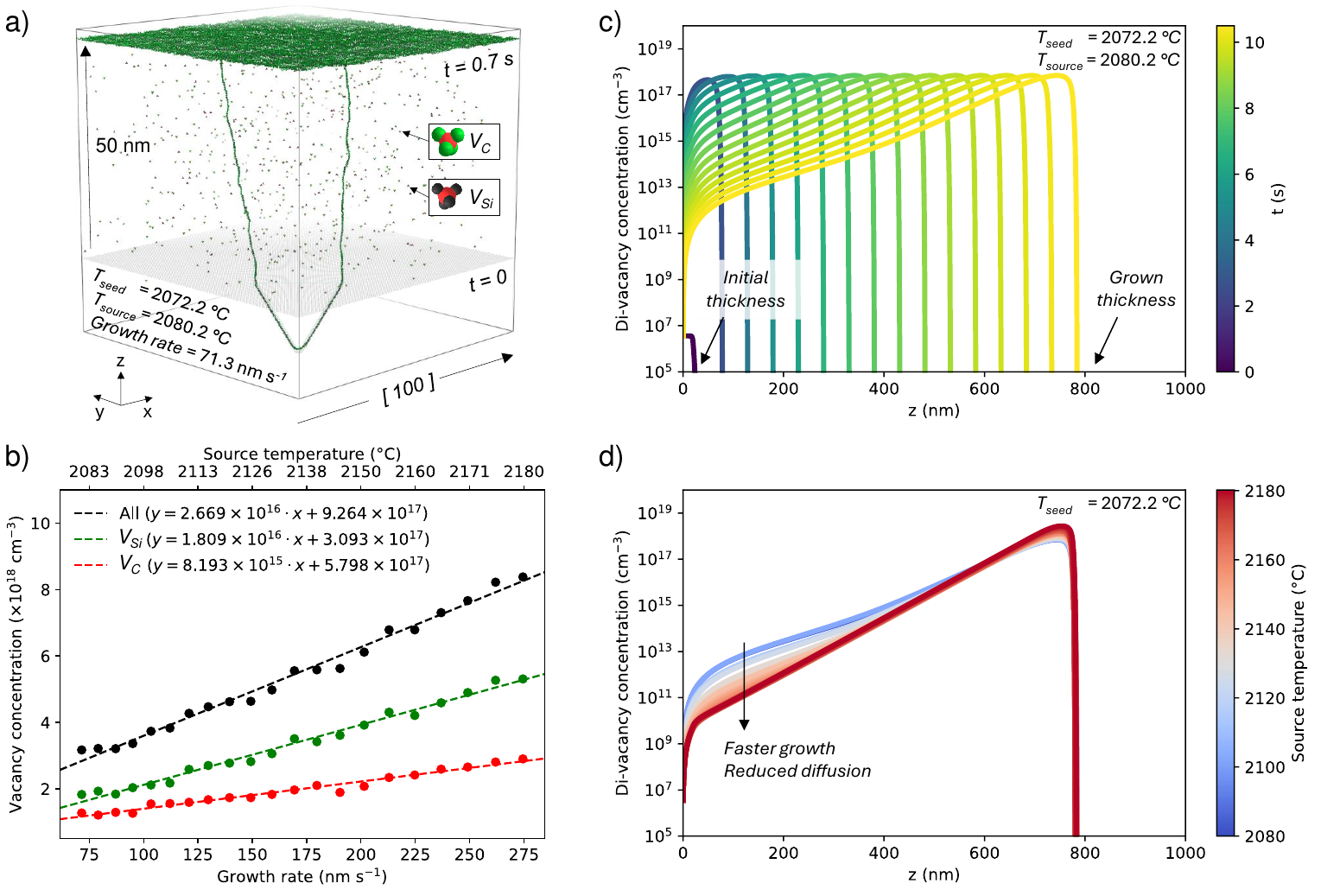}
    \caption{(a) Snapshot after 0.7 s of simulated PVD growth of 3C-SiC layer (50 nm thick film is obtained for the process parameters: $T_{\mathrm{seed}}$=2072.2 °C and $T_{\mathrm{source}}$=2080.2 °C). The simulation initial state (seed) consists of a finite size triple stacking fault defect embedded in \hkl[001] oriented substrate. The atoms (C in black, Si in green)  with coordination deficiency are depicted to show the surface, the evolution of the extended defect (with this convention only the defect's border is visible for a triple stacking fault) and the generated Si and C vacancies. (b) Density of C vacancies (red dots), Si vacancies (green dots) and total vacancies (black dots) generated near the growing surface during PVD growth processes fixed $T_{\mathrm{seed}}$=2072.2 °C and variable $T_{\mathrm{source}}$ in the range $2080.2-2180.2$ °C. The predicted growth rates is also indicated. The dashed lines represent linear best fit of the kinetic Monte Carlo prediction as function of the growth rate (indicated as ``x'' in the legend). (c) Density profiles of di-vacancy complexes $V_{\mathrm{C}}-V_{\mathrm{Si}}$ simulated with the diffusion-reaction model for a PVD growth ($T_{\mathrm{seed}}$=2072.2°C and $T_{\mathrm{source}}$=2080.2 °C) of about 800 nm thick film. Color scale indicates the process time corresponding to the different density profiles. (d) Final density profiles of di-vacancy complexes $V_{\mathrm{C}}-V_{\mathrm{Si}}$ simulated with the diffusion-reaction model for a PVD growth ($T_{\mathrm{seed}}$=2072.2°C and variable $T_{\mathrm{source}}$ in the range 2080.2-2180.2 °C) of about 800 nm thick film. Color scale indicates the value of $T_{\mathrm{source}}$ related to the different density profiles.}
    \label{fig:vacancy-vs-GR}
\end{center}
\end{figure*}

We can simulate the epitaxial growth process on flat and 3D-structured 3C-SiC substrates by exploiting a derived version of the MulSKIPS tool \cite{MulSKIPS} for kinetic Monte Carlo on super-lattice (see Methods). The code is formulated within a specific super-lattice kinetic Monte Carlo approach dedicated to group IV alloys and compounds (or more specifically to crystalline materials with tetrahedral bond symmetry), and it is able to consider the generation and the near surface evolution of all the point defects with nearest neighbor coordination $\leq 4$ in SiC: vacancies ($V_{\mathrm{C}}$, $V_{\mathrm{Si}}$, carbon and silicon lattice site), anti-sites ($A_{\mathrm{C}}$, $ A_{\mathrm{Si}}$ carbon and silicon lattice site), anti-sites-vacancies (Silicon types SAVs, Carbon type CAVs) and the polytype instability leading the formation of stacking faults \cite{fisicaro2020genesis}. We notice that the possible applications of the code for the prediction of the generation efficiency during a realistic growth process with variable process conditions have barely been considered so far. 

An example of simulation code application under realistic conditions is shown in Figure \ref{fig:vacancy-vs-GR} (a), reporting the snapshot after 0.7 s of a simulated physical vapor deposition (PVD) growth process of a 50 thick layer of 3C-SiC, obtained by sublimating SiC powders at $T_{\mathrm{source}}$=2080.2 °C over a substrate at $T_{\mathrm{seed}}$=2072.2 °C. The simulation starts from a finite-size triple stacking fault defect embedded in an otherwise ideal \hkl[001] oriented substrate. In addition to the grown film kinetics and the surface morphology, the simulated generation of vacancies, due to non-ideal local surface reconstruction, and the evolution of the extended defects can be appreciated. 

The near-surface density values of $V_{\mathrm{C}}$ and $V_{\mathrm{Si}}$ generated in the simulated PVD process of 3C-SiC with fixed crystal seed (substrate) temperature $T_{\mathrm{seed}}$=2072.2°C varying the powder source temperature $T_{\mathrm{source}}$ \cite{calogero_crystal2022} are shown in Figure \ref{fig:vacancy-vs-GR} (b). In these simulations ideal substrates are considered as initial states. Anyhow, the inclusion of extended defects in the initial state (as the case shown in Figure \ref{fig:vacancy-vs-GR} (a)) does not lead to significantly different results (compare Tables S1 and S2). The simulations predict a monotonic increase in the crystal seed growth rate with $T_{\mathrm{source}}$ as a consequence of the unbalance between the impingement and sublimation fluxes at the seed surface and a concurrent increase in the efficiency of the vacancy generation. We observe a slightly higher concentration of $V_{\mathrm{Si}}$  compared to $V_{\mathrm{C}}$  within the grown layer. We attribute this to the non-equilibrium atomic kinetics in proximity of the evolving surface, dominated by local deposition event probabilities, which are carefully calibrated using ab-initio calculations \cite{fisicaro2020genesis}. We notice that the growth rate dependence predicted by the simulation is consistent with the experimental trends of the PVD growth processes \cite{calogero_crystal2022}.  

The point defects generated during the growth process at high temperatures evolve in the crystal bulk regions (diffusion, recombination, and clustering phenomena), and this kinetics is not currently included in the Monte Carlo method. Nevertheless, we can use the Monte Carlo predictions of the growth rate and the generation efficiency in a diffusion-reaction framework modeling the bulk evolution. The model and the preliminary calibration of the related parameters are discussed in Section S1.3. It consists of coupled diffusion-reaction equations for the $V_{\mathrm{C}}$ and $V_{\mathrm{Si}}$ defects generated during the growth and the $V_{\mathrm{C}}-V_{\mathrm{Si}}$ complex that forms upon the encounter of the two defects. Along with the complex formation, the reactive terms include the recombination contributions with interstitial carbon and silicon atoms $I_{\mathrm{C}}$ and $I_{\mathrm{Si}}$ whose density is assumed to remain constant at the thermodynamic equilibrium level during the process. The crystal growth and the generation efficiency are modeled in this continuum framework within a phase field formalism similarly to the method introduced in Ref. \cite{lamagnaPhysRevB2007}, ruled by parameters calibrated with the results of Figure \ref{fig:vacancy-vs-GR} (b).             

The MulSKIPS generation kinetics can be directly mapped in the diffusion reaction framework with a phase field equation which, solved as a stand-alone equation, recovers the time dependent surface motion and the concurrent generation of the constant $V_{\mathrm{C}}$ and $V_{\mathrm{Si}}$ density in the evolving solid region (i.e. the average depth dependent equivalent profile of the atomistic one shown in Figure \ref{fig:vacancy-vs-GR} (a)). The coupling of this equation with the diffusion-reaction model allows the recovery of more reliable residual damage predictions where OVPDs reside mainly in stable di-vacancy complexes. The time-dependent density profiles of di-vacancy complexes $V_{\mathrm{C}}-V_{\mathrm{Si}}$ simulated with the diffusion-reaction model after the PVD growth of about 800 nm thick film for the process parameter $T_{\mathrm{seed}}$=2072.2 °C and $T_{\mathrm{source}}$=2080.2 °C are shown in Figure \ref{fig:vacancy-vs-GR} (c). The final profile is frozen at the end of the growth, but it could evolve further in real processes during the finite time required for quenching. The evolution shown in  Figure\ref{fig:vacancy-vs-GR} (c) is the result of thermally activated competitive formation-recombination phenomena and defect diffusion which cause a peak of di-vacancy density close to the moving surface and a relatively fast reduction of the complex concentration in the bulk region. We notice that average values of the residual single-vacancy densities are significantly lower and the peak densities for $V_{\mathrm{C}}$ and $V_{\mathrm{Si}}$ are roughly half of the peak density of $V_{\mathrm{C}}-V_{\mathrm{Si}}$.
The impact of $T_{\mathrm{source}}$ variation on the $V_{\mathrm{C}}-V_{\mathrm{Si}}$ profiles, obtained for films grown with the same thickness (800 nm), can be appreciated in Figure \ref{fig:vacancy-vs-GR} (d). An increase in $T_{\mathrm{source}}$ for constant $T_{\mathrm{seed}}$ yields faster growth and higher densities of generated defects. The combination of these two trends leads to a larger peak density and a lower bulk average density in the $V_{\mathrm{C}}-V_{\mathrm{Si}}$ profiles, where the latter is due to the reduced total diffusion time.   

The scenario shown in Figures \ref{fig:vacancy-vs-GR} could represent a typical outcome of the process simulation in a QT-CAD study: verifiable predictions of material modifications, in terms of spacial distributions of active atomic centers of interest for the quantum application under investigation, correlated to the real process parameters (in this case $T_{\mathrm{seed}}$ and $T_{\mathrm{source}}$). The study is usually motivated by the quests for an optimization of the process parameters or for the definition of a proper process window or for the exploration "in silico" of alternative solutions (for example, a proper time dependence of the process parameters). This use is rather common for the conventional semiconductor technology, although the counterpart of widely applied models with a high degree of accuracy in the case of elemental semiconductors (i.e. Si, Ge) barely exists for alloys like SiC. 
To conclude this section, we note that the continuum evolution model, whose results are discussed in Figures \ref{fig:vacancy-vs-GR} (c,d) can be generalized to simulate point defects' generation by means of energetic beams, through an appropriate definition of the heat source used for thermal annealing and of the chemical reactions' terms in the core partial derivative equations. Of course, in these cases explicit solutions for all defect types in SiC are necessary. We further note that we have focused on the global defects' kinetics and diffusion, regardless of their charge state. This is reasonable, given the intrinsic nature of the semiconductor and the high processing temperatures considered in this study, which would produce a near balance in positive and negative defect concentrations, along with significant thermal broadening \cite{RevModPhys.61.289}. However, it should be noted that both the KMC \cite{Raciti20231MSSP} and diffusive models can be generalized for moderately and highly doped materials. 



\begin{figure*}
    \includegraphics[width=1.0\textwidth]{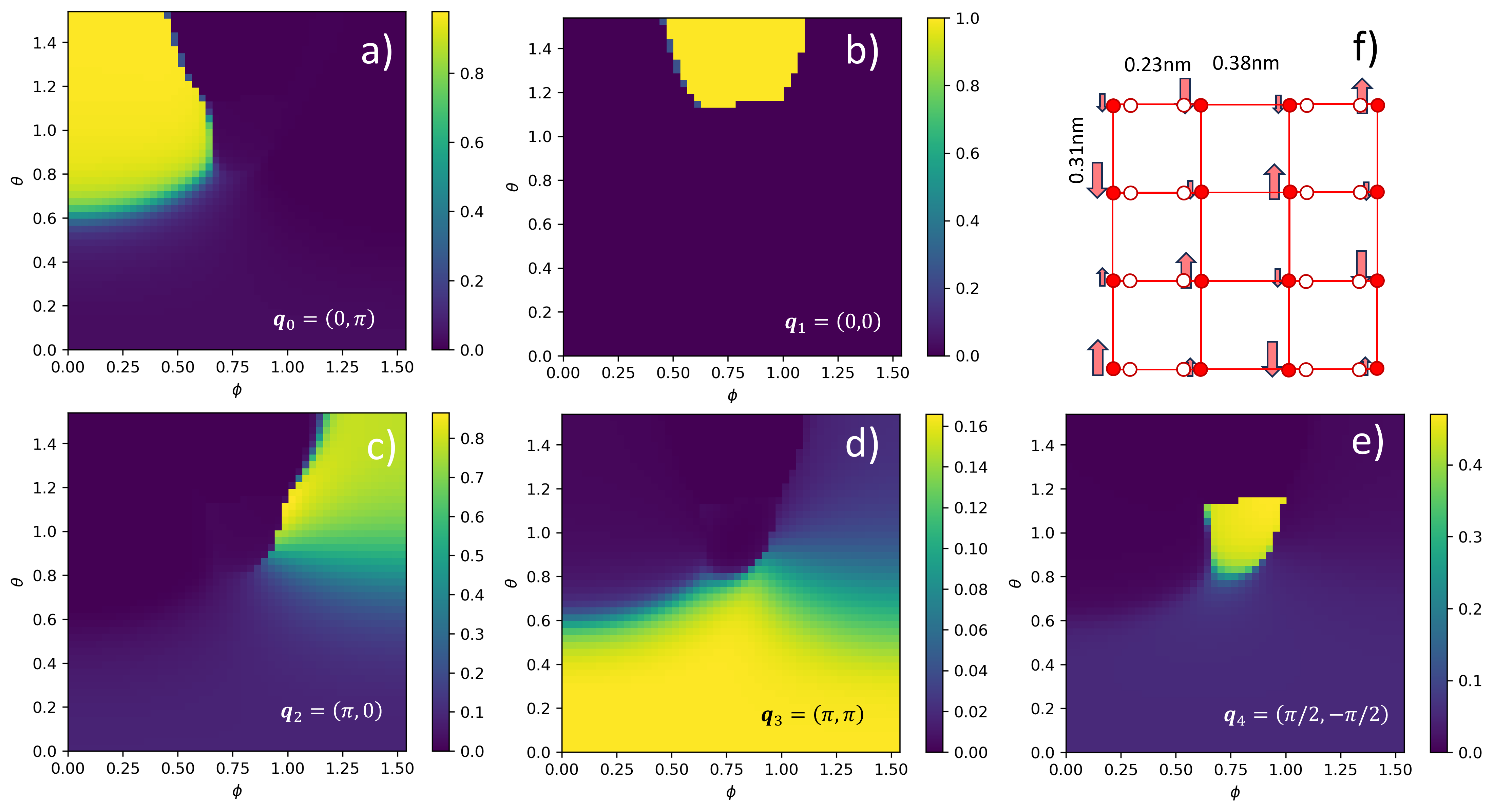}
    \caption{a-e) Maps, as a function of the external $B$ field direction in polar coordinates $(B,\theta,\phi)$, of the structure factors $S(\mathbf{q})$ calculated with the modified $\Delta/2$ XXZ model for the interacting H nuclear spins on a ($2\times 1$) reconstructed \hkl(001) 3C-SiC surface for the 5 different spin $\mathbf{q} \equiv (q_x,q_y)$ waves indicated in the panels. (a) $S(\mathbf{q}_0$) with $\mathbf{q}_0 = (0, \pi)$ structure factor the ferromagnetic (along X direction) and antiferromagnetic order (along Y direction). (b) $S(\mathbf{q}_1$) with $\mathbf{q}_1 = (0, 0)$ structure factor for the ferromagnetic order. (c) $S(\mathbf{q}_2$) with $\mathbf{q}_2 = (\pi, 0)$ structure factor for the ferromagnetic (along Y direction) and antiferromagnetic order (along X direction). (d) $S(\mathbf{q}_3$) with $\mathbf{q}_3 = (\pi, \pi)$ structure factor for the antiferromagnetic order. (e)  $S(\mathbf{q}_4$) with $\mathbf{q}_4 = (\pi/2,-\pi/2)$) structure factor for the diagonal ferromagnetic-antiferromagnetic chains. (f) Pictorial representation of the $(\pi/2, \pi/2)$ spin wave in the H sites present in the surface.}
    \label{fig:SqMAPS100recon}
\end{figure*}

\begin{figure*}
\begin{center}
    \includegraphics[width=0.75\textwidth]{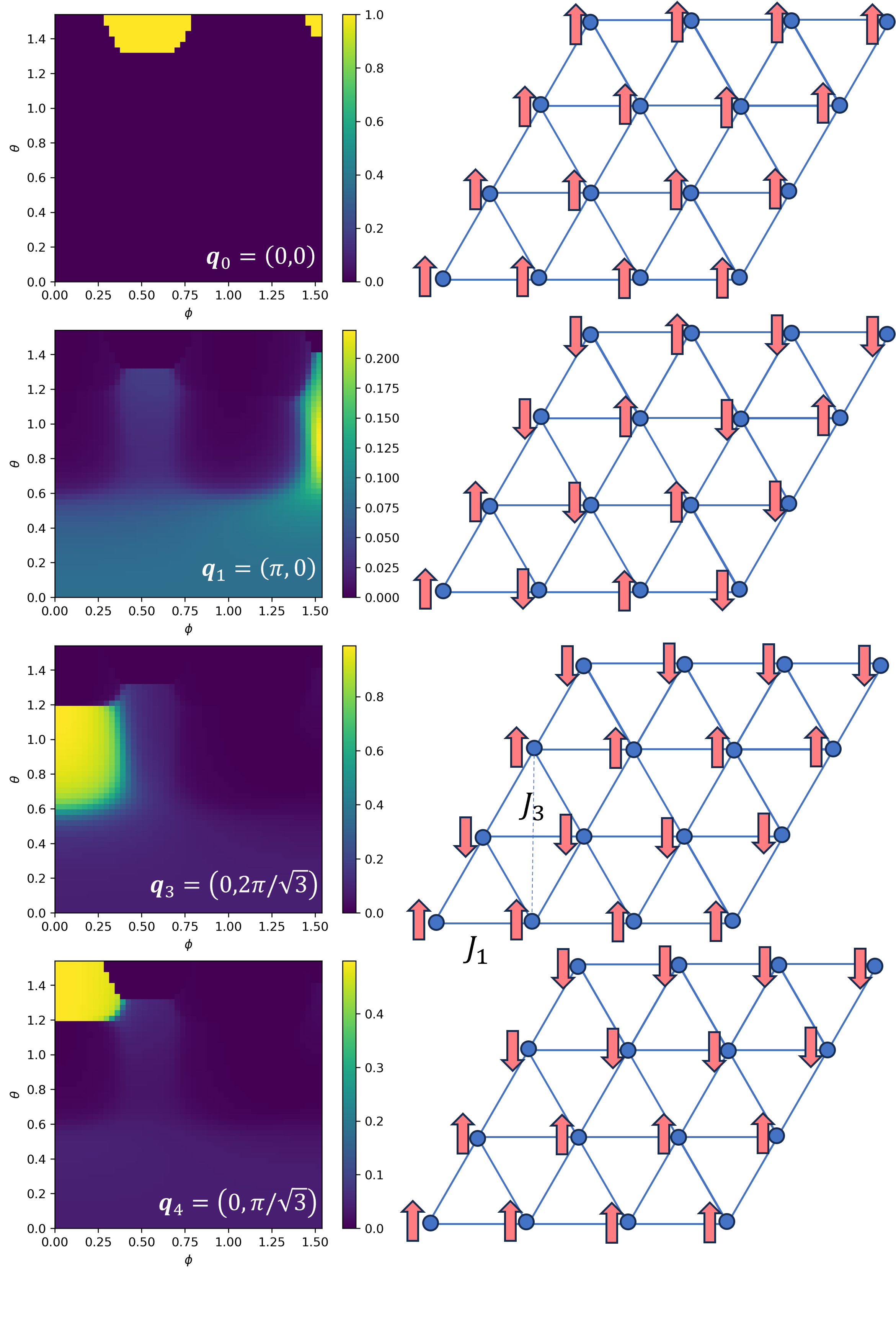}
    \caption{Left hand panels) maps, as a function of the external $B$ field direction in polar coordinates $(B,\theta,\phi)$, of the structure factors $S(\mathbf{q})$  calculated with the modified $\Delta/2$ XXZ model for the interacting H nuclear spins on \hkl(111) 3C-SiC surface for the 4 different spin $\mathbf{q} \equiv (q_x,q_y)$ waves indicated in the panels with at ferromagnetic order in one of the two cartesian direction. In the right hand panels the corresponding pictorial representations of the spin waves are shown. }
    \label{fig:SqMAPS111_1}
\end{center}
\end{figure*}

\begin{figure*}
\begin{center}
    \includegraphics[width=0.8\textwidth]{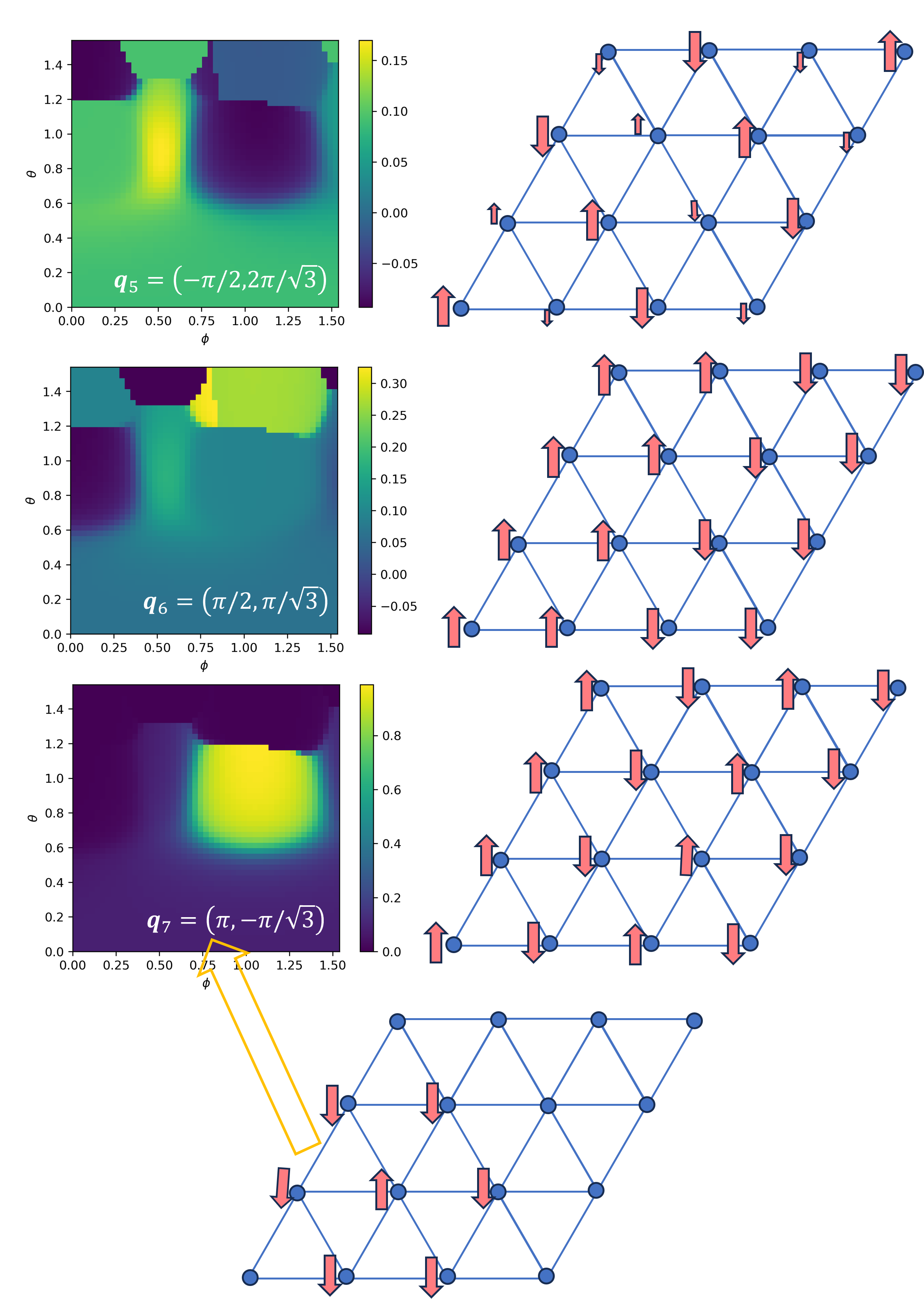}
    \caption{Left hand panels) Maps, as a function of the external $B$ field direction in polar coordinates $(B,\theta,\phi)$, of the structure factors $S(\mathbf{q})$ calculated with the modified $\Delta/2$ XXZ model for the interacting H nuclear spins on \hkl(111) 3C-SiC surface for the 4 different spin $\mathbf{q} \equiv (q_x,q_y)$ waves indicated in the panels. Here the case with both non-zero $(q_x,q_y)$ values are shown. In the right hand panels the corresponding pictorial representations of the spin waves are shown. In the bottom panel a pictorial representation of the state with short range AF order, which characterizes the low $\phi$ values, is also shown}
    \label{fig:SqMAPS111_2}
\end{center}
\end{figure*}

\section{\label{sec:spinwaves} Instabilities of H nuclear spins ordered phase}

In the previous section we highlighted that different hydrogenation processes allow for the formation of large flat regions ({\textmu}m wide) where H atoms are distributed in a quasi-2D lattice configuration on low Miller index 3C-SiC surfaces. The nuclear spin associated to these quasi 2D lattices in the presence of relatively high magnetic fields should be governed by the quantum XXZ spin model in Eq. \ref{eq:XXZ} with variable spin-spin interactions, which could manifest a rich scenario of magnetic phases depending on the direction of the external magnetic field. We notice that similar systems of interacting point-planar spins (specifically nitrogen-vacancy spins in diamond and surfaces with mono-layer fluoride coverage) have been indicated as a possible quantum simulator for the direct study of frustrated quantum magnetism in systems implementing quantum Hamiltonians \cite{cai2013natureQS}. In this section we discuss this issue. We characterize the ground state spin order and eventual frustrated quantum magnetism by means of exact calculations on ($n\times n$) clusters with periodic boundary conditions. Python APIs based on the QuTiP Toolbox have been implemented for the scope \cite{QuTiP}.      

As indicated already, the spin-spin coupling parameter in Eq. \ref{eq:XXZ} depends on the space positioning of the H-related spins and the direction of the B field with respect to the X-Y plane where the atoms reside. In section S2.2, the competing character of anti-ferromagnetic and ferromagnetic interaction is studied in detail for the perfect square lattice arrangement of the ideal \hkl(001) 3C-SiC surface case. In this section we study the system with the Eq. \ref{eq:XXZ} while implicitly assuming the reliability of an adiabatic preparation procedure from the regime of the validity of RWA in a disordered state to the ground state. In the next section we will critically discuss the time protocols in the presence of the perturbation of the near-to-surface defects. The spin order is analyzed in terms of the structure factor $S(\mathbf{q})$ defined as 
\begin{equation}
    S(\mathbf{q}) = \frac{1}{N^2} \left< \sum_{i,j} e^{i \mathbf{q} \cdot \left( \mathbf{r_i} - \mathbf{r_j} \right)} \sigma_z(i) \sigma_z(j)  \right> 
\end{equation}
In Figure \ref{fig:SqMAPS100recon} (a-e) we show the maps of the structure factors $S(\mathbf{q}_i, i=0,4)$ calculated in a $(4\times 4)$ periodic cluster for varying direction of the external (driving) magnetic field and for the five symmetries of spin-spin waves which show significant stability regions. The latter are defined as areas (in polar coordinates of the magnetic field) where spin waves, defined by the vector q, are most likely to be observed. The structural factor S(q) quantifies the probability of detecting nuclear hydrogen spin waves as the magnetic field direction varies. In Figure \ref{fig:SqMAPS100recon} (f) the spin arrangement for the diagonal ferromagnetic-antiferromagnetic chains spin order is graphically depicted, with reference to dimerized configuration of H atoms.   

We can notice that the X-Y symmetric scenario of the square lattice discussed in section S2, is modified by the presence of the dimerized state. The regions dominated by the ferromagnetic $S(\mathbf{q}_1$) with $\mathbf{q}_1 = (0,0)$ (Figure \ref{fig:SqMAPS100recon}(b)), antiferromagnetic $S(\mathbf{q}_3$) with $\mathbf{q}_3 = (\pi, \pi)$ (Figure \ref{fig:SqMAPS100recon}(d)) and diagonal ferromagnetic-antiferromagnetic chains $S(\mathbf{q}_4$) with $\mathbf{q}_4 = (\pi/2,-\pi/2)$ (Figure \ref{fig:SqMAPS100recon}(e)) are strongly distorted with respect the corresponding X-Y symmetric ones in Figure S5. The symmetry breaking has a stronger impact for the ferromagnetic-antiferromagnetic order, whose stability is strongly suppressed when the dimerization direction is parallel to the direction of the anti-ferromagnetic spin arrangement (compare panels (c) in Figure \ref{fig:SqMAPS100recon} and Figure S5).        

The triangular symmetry of the 2D lattice which forms on the \hkl(111) 3C-SiC surface leads to a significantly richer scenario for the H nuclear spin wave instabilities.
In Figures \ref{fig:SqMAPS111_1} and \ref{fig:SqMAPS111_2} we show the maps of the structure factors $S(\mathbf{q}_i, i=0,6)$ calculated in a $4\times 4$ periodic cluster for varying direction of the external (driving) magnetic field and for the seven symmetries of spin-spin waves which show significant stability regions in this triangular lattice. To enhance the understanding, for each map the corresponding arrangement of spins is shown. We notice that the symmetries with $q_x=0$ or $q_y=0$ (figure \ref{fig:SqMAPS111_1}) can be stabilized for planar changes of the external field direction ($B_x=0$ or $B_y=0$), while the symmetries with both non-zero $q_x$ or $q_y$ values can only be stabilized with 3D rotations of $\mathbf{B}$. It's worth observing the absence of fully anti-ferromagnetic state, which instead dominates the low $\theta$ region for both the ideal square and reconstructed square lattices (Figures S5 and \ref{fig:SqMAPS100recon}). The geometric frustration of the triangular lattice hinders the long-range anti-ferromagnetic order leading to the emergence of local arrangement of next-neighbors spins with opposite polarity as the one pictorially shown in the low-most panel of Figure \ref{fig:SqMAPS111_2}.  The rich scenario is promoted also by the impact of the third neighbors interaction which, in consideration of the shorter distances and the external field direction, can be stronger and opposite in sign with respect of the first and second neighbor coupling parameters (see Ref. \cite{cai2013natureQS} and also the comments on the coupling parameter maps in Figure S6 within section S3).

\section{\label{sec:protocols} Time-dependent manipulation of H-defect spins coupled systems} 
Multiple ordered states which can be stabilized in a periodic distribution of two-levels interacting quantum system under controllable external parameter, as the hydrgens/3C-SiC cases discussed in the previous section, are of course the key features of a quantum simulator. We notice that solid-state systems are not the only possibility and relevant studies have been performed on 2D lattices of Rydberg atoms \cite{Chen2023,Browaeys2020,Labuhn2016,Bluvstein2021}. Anyhow, the implementation of these systems in real quantum simulators requires the design and realization of several time-dependent control steps. These include initialization with high fidelity, selective polarization cycles, adiabatic preparation of a close to ground state configuration, read-out, as discussed, \textit{e.g.}, in Ref. \cite{cai2013natureQS} for the nitrogen-vacancy + fluoride spins in diamond. The results discussed in the latter work can be generalized for di-vacancies in the hydrgens/3C-SiC system with proper Hamiltonian parameters' reconfiguration. Of course, the deviation of the real system from the assumed ideal atomic configurations and defect location and, in general, any source of dissipation can critically hinder the successful achievement of the desired protocol results. Here we discuss these issues specifically for the adiabatic quenching protocol which, among others, especially requires careful examination and optimization.           

\begin{figure*}[t]
   \includegraphics[width=1.0 \linewidth]{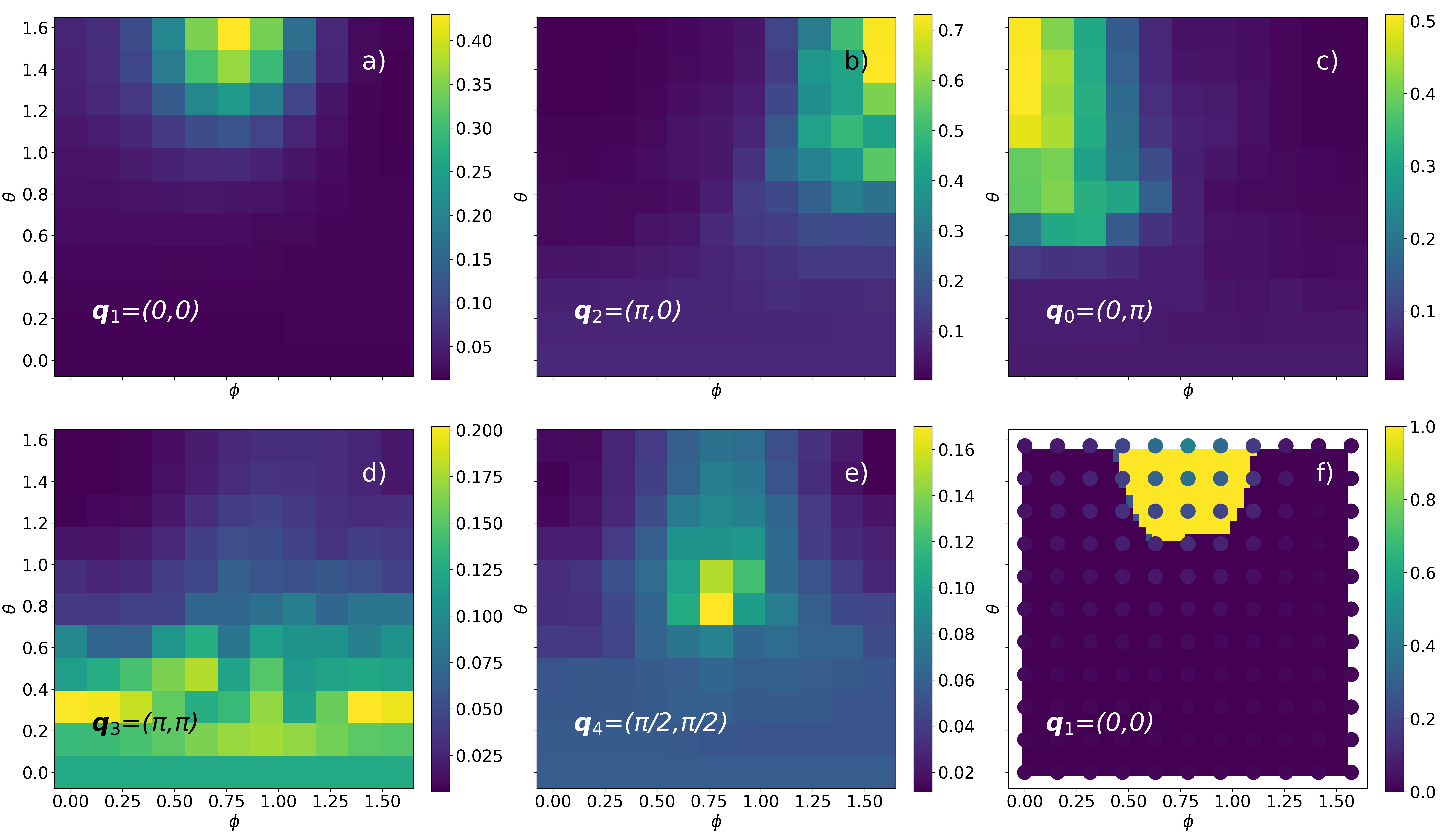}
    \caption{(a-e) Maps, as a function of the external $B$ field direction in polar coordinates $(B,\theta,\phi)$, of the structure factors $S(\mathbf{q})$  calculated with the modified $\Delta/2$ XXZ model after the adiabatic quenching protocol discusses for the interacting H nuclear spins on \hkl(001) 3C-SiC surface. Each panel corresponds to a distinct spin wavevector  $\mathbf{q} \equiv (q_x,q_y)$ , with cases shown where both $q_x$ and  $q_y$ are non-zero.  In panel (f), the values of the structure factor for the ground state are compared with those obtained after the adiabatic quenching protocol (i.e., panel (a)), represented as a scatter plot on the same scale. }
    \label{fig:figure_GS_onltn}
\end{figure*}
In a quantum simulator application, the state preparation in the XXZ effective model is obtained applying a radio-frequency field with amplitude $\Omega_{RF}$ and de-tuning $\Delta_{RF}$ with respect to the resonant hydrogen Larmor frequency $\gamma_H B$. As a consequence, the effective model can be written as
\begin{equation}
\begin{split}
\mathcal{H}_{eff} &= \sum_{i} \Omega_{RF} \sigma_x(i) + \sum_{i} \Delta_{RF} \sigma_z(i) +\frac{\mu_{0}}{4 \pi}\sum_{i,j} \frac{\gamma^n_{i}\gamma^n_{j}}{r_{i,j}^3} \left(1 - 3  (\hat{\mathbf{r}}_{i,j}^B)^2\right) \left[\sigma_z(i) \sigma_z(j) -\Delta \left(\sigma_x(i)\sigma_x(j) + \sigma_y(i)\sigma_y(j) \right)  \right]  \\ &= \sum_{i} \Omega_{RF} \sigma_x(i) + \sum_{i} \Delta_{RF} \sigma_z(i) + \mathcal{H}_{XXZ}. \\
\label{eq:XXZ_eff}
\end{split}
\end{equation}

In the preparation step, the amplitude of the driving RF field is the dominant energy $\Omega_{RF}$ of the effective spin interaction and the ground state can be reliably approximated by the product state $\ket{X} = \bigotimes_{i} \ket{\downarrow_x(i)}$ of the $\sigma_x(i)$ eigenstates $\ket{\downarrow_x(i)} = 1/\sqrt{2} \left[\ket{ \downarrow(i)}-\ket{\uparrow(i)}\right]$. The adiabatic switching procedure aims to stabilize a close-to-ground-state of $\mathcal{H}_{XXZ}$ with good fidelity by a gradual reduction of the RF field amplitude $\Omega_{RF} (t)$. 
The success of the procedure is not in general guaranteed since it can depend on the switching modality and, more fundamentally, from the initial state and the target ground state which should be obtained with good approximation after the time dependent dynamics. We can simulate the adiabatic preparation using a realistic functional dependency of $\Omega_{RF} (t)$ and in general the optimal choice can also be numerically evaluated. 
We have simulated the evolution of the model $\mathcal{H}_{eff}$ for the case of interacting H nuclear spins on a ($2\times 1$) reconstructed \hkl(001) 3C-SiC surface and variable directions of the external field. We consider a quadratic decrease to zero of $\Omega_{RF} \left( t \right)= \Omega_{RF}(0)\cdot \left( 1 - t/\tau_1 \right)^2$ for $t < \tau_1 $, followed by a free evolution with $\Omega_{RF} = 0$ for $\tau_1 \leq t \leq \tau_2 $. In Figure \ref{fig:figure_GS_onltn} we show the structure factor maps for different spin waves for $\Omega_{RF}(0) = 200$ MHz,  $\tau_2 = 2  \tau_1 = 5$ ms and negligible ($\sim 0$) value of de-tuning. Comparing the maps in Figures \ref{fig:figure_GS_onltn} and \ref{fig:SqMAPS100recon} we can conclude that the reliability of the adiabatic preparation could strongly depend on the symmetry of the target ground state. Indeed, for the antiferromagnetic (see \ref{fig:figure_GS_onltn} d)) and for the mixed ferromagnetic-antiferromagnetic (see \ref{fig:figure_GS_onltn} b),c) and e)) symmetries the structure factor values are close to the ground state one, while for the ferromagnetic symmetry the values of $S(\mathbf{0,0})$ are significantly lower than the ground state one (see also direct comparison in Figure \ref{fig:figure_GS_onltn} f) ). The transition among different symmetries is smoother for the dynamically stabilized state, although the phase stability scenario in view of a quantum simulator application is globally recovered with sufficient fidelity already with this quenching procedure of the RF field, in spite of the reduced fidelity for the ferromagnetic sector. 

\begin{figure*}[t]
    \includegraphics[width=1.0 \linewidth]{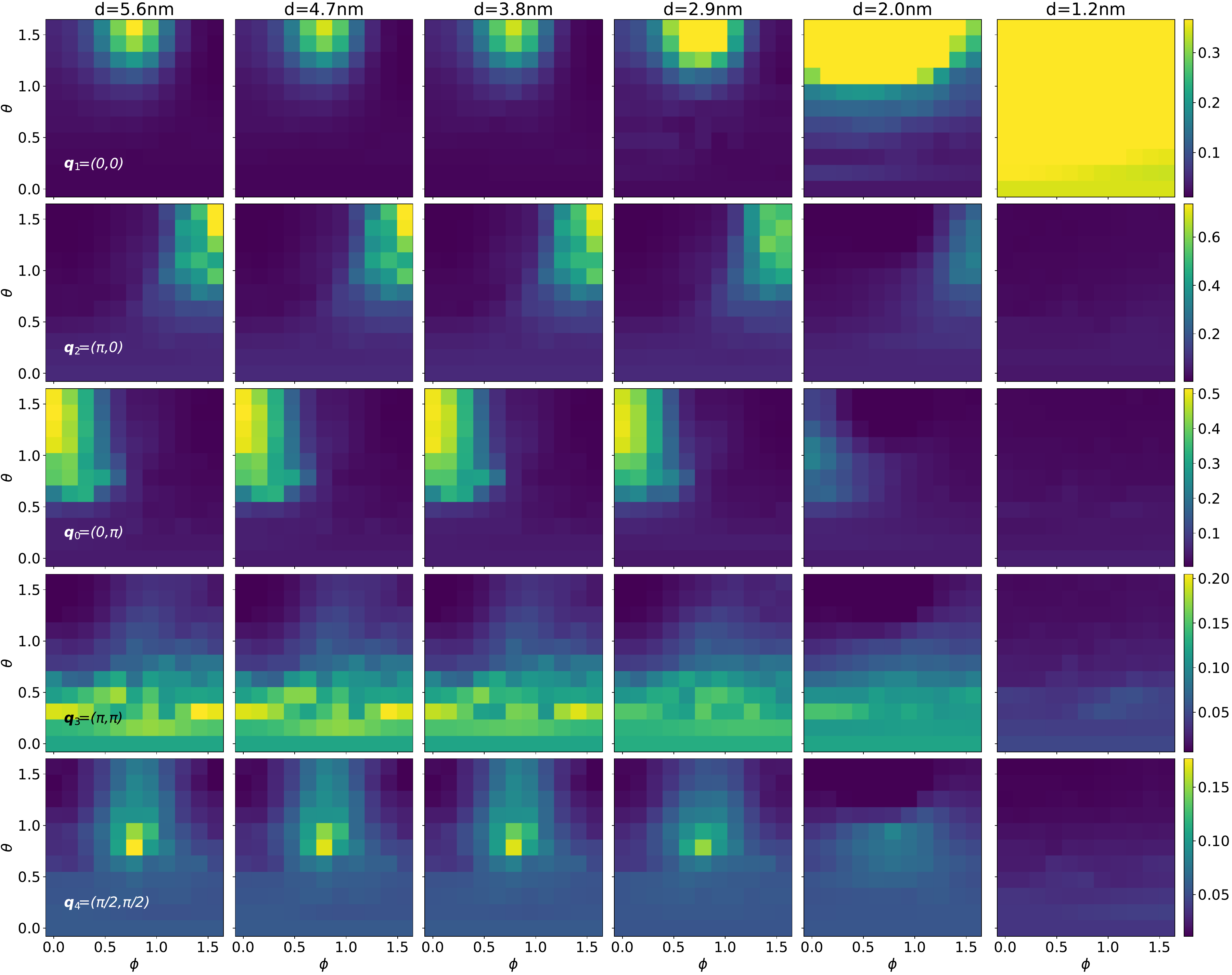}
    \caption{ Maps, as a function of the external $B$ field direction in polar coordinates $(B,\theta,\phi)$, of the structure factors $S(\mathbf{q})$  calculated with the modified $\Delta/2$ XXZ model after a the adiabatic quenching protocol discusses for the interacting H nuclear and di-vacancies electron spins on \hkl(001) 3C-SiC surface for the 5 different spin waves indicated as $\mathbf{q} \equiv (q_x,q_y)$ in each row. Panels from left to right show the results for decreasing distance between defect and the plane where the H nuclei reside.}
    \label{fig:figure_dist_mapnn}
\end{figure*}

Weak dipole-dipole coupling between bulk di-vacancies and nuclear spins makes feasible the polarization transfer between the di-vacancy spin and the nuclear spin (i.e. the preparation of the initial state before the adiabatic preparation) and the read-out of the nuclear spin state after the procedure when the di-vacancy acts as a probe for the measurements \cite{cai2013natureQS}. However, the di-vacancy's presence in the proximity of the surface could alter the adiabatic procedure, hindering the ideally estimated fidelity level of the final state (Figure \ref{fig:figure_GS_onltn}).   \\
We have evaluated the impact on the adiabatic procedure of the interaction with the "closest" di-vacancy for a periodic cluster model of composite di-vacancy plus H nuclear spins. In our modelling, for the sake of simplicity, we consider a di-vacancy transversely oriented with respect to the (001) surface, in line with recent studies \cite{zhu_galliananolett2024}. As for the previous analysis we have considered the ($2\times 1$) reconstructed \hkl(001) 3C-SiC surface. We assume that the di-vacancy state is that emerging from previous manipulation steps, \textit{i.e.,} initialization with an off-resonant laser and micro-wave driving at the Rabi amplitude $\Omega_{VV}$ on resonance with the lowest spin transition at a given value of the uniform magnetic field $B$. Within the same approximation level due to the large external field the spin flip transition of the vacancy spin are inhibited in the effective coupled model and the electron spin nuclear spin dipole dipole coupling is approximated as 
\begin{equation}
\mathcal{H}_{VV-H} = \frac{\Omega_{VV}}{2} \sigma^{VV}_{x} + \frac{1}{2} \left ( \mathbf{I} + \sigma^{VV}_{z} \right) \cdot \sum_{i} \left[A_{z,z}(i) \sigma_z(i) + A_{z,x}(i) \sigma_x(i) + A_{z,y}(i) \sigma_y(i)\right] +  \mathcal{H}_{eff} 
\label{eq:full}
\end{equation}
where $\sigma^{VV}_{x,y,z}$ are the effective operators coupling the two lowest spin levels and $A_{\alpha,\beta}(i)$ are the components of the electronic-nuclear dipole-dipole tensor for the nuclear spin $i$.

We have analyzed such an adiabatic preparation dynamics in the presence of the defect with the model of Eq. \ref{eq:full} for an initial state $\ket{X} \otimes 1/\sqrt{2} \left[\downarrow (VV)-\uparrow (VV) \right] $, varying the distance between the di-vacancy center and the plane where the H atoms reside from 5.6 to 1.2 nm. The point dipole approximation is assumed to estimate the dipole-dipole tensor components.
The results are shown as structure factor maps in Figure \ref{fig:figure_dist_mapnn}. \\
At sufficiently large distance $d=5.6$ nm the interaction with the defect does not alter the final-state fidelity significantly and the structure factor maps do not show important differences with respect to those obtained in the non-interacting case. During and after the RF amplitude reduction procedure the defect state remains stuck at the initial one (see section S4) and the dipole-dipole interaction only slightly perturbs the H spin final ordered state. We notice that the polarization transfer and read-out procedures are feasible using a defect at this distance as a probe, therefore the quantum simulation global protocol should be safe in this case. \\
The antiferromagnetic and mixed spin correlations are gradually weaken as the defect approaches the surface (see Figures
\ref{fig:figure_dist_mapnn}) while a non-univocal behavior appears for the ferromagnetic correlation $\mathbf{q}_1 \equiv (0,0)$ , which decreases for distances in the range $d=5.6-3.8$ nm and then increases for lower distances, becoming totally dominant for $d=1.2$ nm. For the latest case, a strong polarization of the spin defect in the $z$ direction characterizes the end of the adiabatic procedure (see also section S4). We should notice that average values in the order of a few tens of kHz trigger dipole-dipole interactions between defect and nuclear spins in the computational cluster for the lowest distance case. For such high values of dipole-dipole coupling, this interaction dominates with respect to the other terms. Moreover, also the accuracy of some of the approximations could be questionable. However, in spite of these considerations, the main conclusion is the necessity of an accurate analysis of the protocol results with support of the theoretical simulation to consider the impact of the space distribution of the defects when quantum simulators or other protocols are under investigation.

\section{Conclusion}
The computational research presented in this paper aims to provide a complete study of quantum design for a system of potential interest in various quantum technologies, from sensing to simulators.  Active quantum objects are the electronic spins associated with OVPDs (in particular $V_C V_{Si}$ di-vacancies) in 3C-SiC coupled with nuclear spins of H-type functionalization of the 3C-SiC surface. The atomic nature of the centers, in general, requires an atomistic resolution for the predictions/descriptions in the overall modeling approaches. 
A particular, but not unique, focus is given here to solid-state quantum simulator applications of the studied system. Therefore, a detailed study of the target correlated quantum-state objective of the simulation (i.e. the spin instabilities of a  $\Delta/2$ XXZ model with variable coupling parameters) have been discussed for two possible 2D lattice distributions of the H-related spins, corresponding to two low Miller index \hkl(001) and \hkl(111) surface reconstructions.  

The richness of ordered phases that can be initialized and probed with weakly interacting electronic spins, along with the almost natural access to both the \hkl(001) and \hkl(111) surfaces (e.g. as a consequence of hetero-epitaxial growth on corresponding Si oriented substrates) makes the 3C-SiC polytype particularly interesting for such quantum operations. We notice that, with the necessary differences, the results for the \hkl(111) case can be extended to the \hkl(0001) polar surface of the hexagonal polytypes, whilst non-polar surfaces of these polytypes cannot recover the atomic order of the \hkl(001) cubic surface.    

The accurate atomistic simulations of the synthesis and manipulation processes used for material modifications necessary to fabricate a quantum device/prototype could provide crucial support to its design, suggesting alternative solutions or enabling a strongly enhanced control over the techniques. In Section \ref{sec:process}, we explore the significance of these tools and propose combining SlKMC simulations with a continuum evolution model to determine the conditions for controlled OVPDs. These conditions would generate varying distributions of active $V_C V_{Si}$ centers in as-grown 3C-SiC substrates under fast, non-equilibrium physical vapor deposition growths.
The simulations discussed here consider fixed process parameters, but time-dependent process parameters can also be implemented based on the specific scope of the optimal process design (e.g. to obtain near-surface zones enriched or depleted of OVPDs). Another observation from the analysis of predicted OVPD profiles is that a significant density of OVPDs could be already be present at the surface proximity of the 3C-SiC substrates before any subsequent defect production process (e.g. implantation or irradiation).   

The latter results could have a strong impact in the quasi ground state stabilization of nuclear spins through the adiabatic quenching step of the quantum simulator protocols. Indeed, this step is highly critical among the quantum simulator sequence of steps (i.e. $\ket{0}$ defect spin  initialization by off resonant laser irradiation, $\ket{0}$, $\ket{+}$ mixing with MW field to obtain Rabi-like effective dynamics in the RF range, resonant RF manipulation for independent spin initialization, quenching of RF amplitude for quasi-GS generation, read-out of the quasi-GS again with resonant RF manipulation \cite{cai2013natureQS} ); since also for the pure nuclear spin model we calculated that the relative fidelity after quenching, with respect to the many spin ground states, strongly depends on the direction the external magnetic field and the corresponding spin order. Moreover, for particular symmetries (e.g. the ferromagnetic one for the \hkl(001) H-spins 2D lattice) the order is significantly weaker with respect to the one in the true GS.     

These findings should not totally hinder the quantum simulator operations which can, in principle, also be supported by optimal control strategies for the adiabatic quenching step. Anyhow, the requirement of weak interactions between interrogating nuclear spins and probing defect spins must be satisfied. Indeed, the proximity of the defect (at a distance below 3 nm from the surface where the H atoms reside) significantly alters the structure factor maps and makes the overall protocols not applicable. An accurate material process where a few nm region fully depleted by OVPDs is necessary and epitaxial growth simulations indicate that this crystal quasi-ideality is achievable by means of quasi-equilibrium slow growth at the end of the process. 

A near-surface defect-rich material could be interesting for entanglement-enhanced sensing applications, since strong hyperfine coupling between electron and nuclear spin could allow for the extension of the protocols in ref. \cite{zaiser2016sensor} for the bulk NV center in diamond. The surface proximity could, of course, open new perspectives for localized measurements not feasible with bulk sensing centers and for electrical read-out modality \cite{McCallumPhysRevLett.132.146902}. We highlight again the importance of QT-CAD tools (from material process simulations to the accurate quantum dynamics study of the protocols) to support the experimental realization and characterization of this or other complex quantum devices.          

\medskip
\textbf{Supporting Information} \par 
Supporting Information is available from the Wiley Online Library or from the author.

\medskip
\textbf{Acknowledgements} \par 
The work has been partially funded by the Italian Ministry University and Research (MUR) in the framework of project PNRR Partenariato PE4 NQSTI Quantum, Grant No. PE0000023. I.D and G.F. acknowledge the funding of  Ministero delle Imprese e del Made in Italy (MIMIT) under IPCEI Microelettronica 2, MicroTech for Green project 

\medskip

%
\bibliographystyle{MSP}
\bibliography{manuscript}

\begin{figure}
\textbf{Table of Contents}\\
\medskip
  \includegraphics[width=1.0 \linewidth]{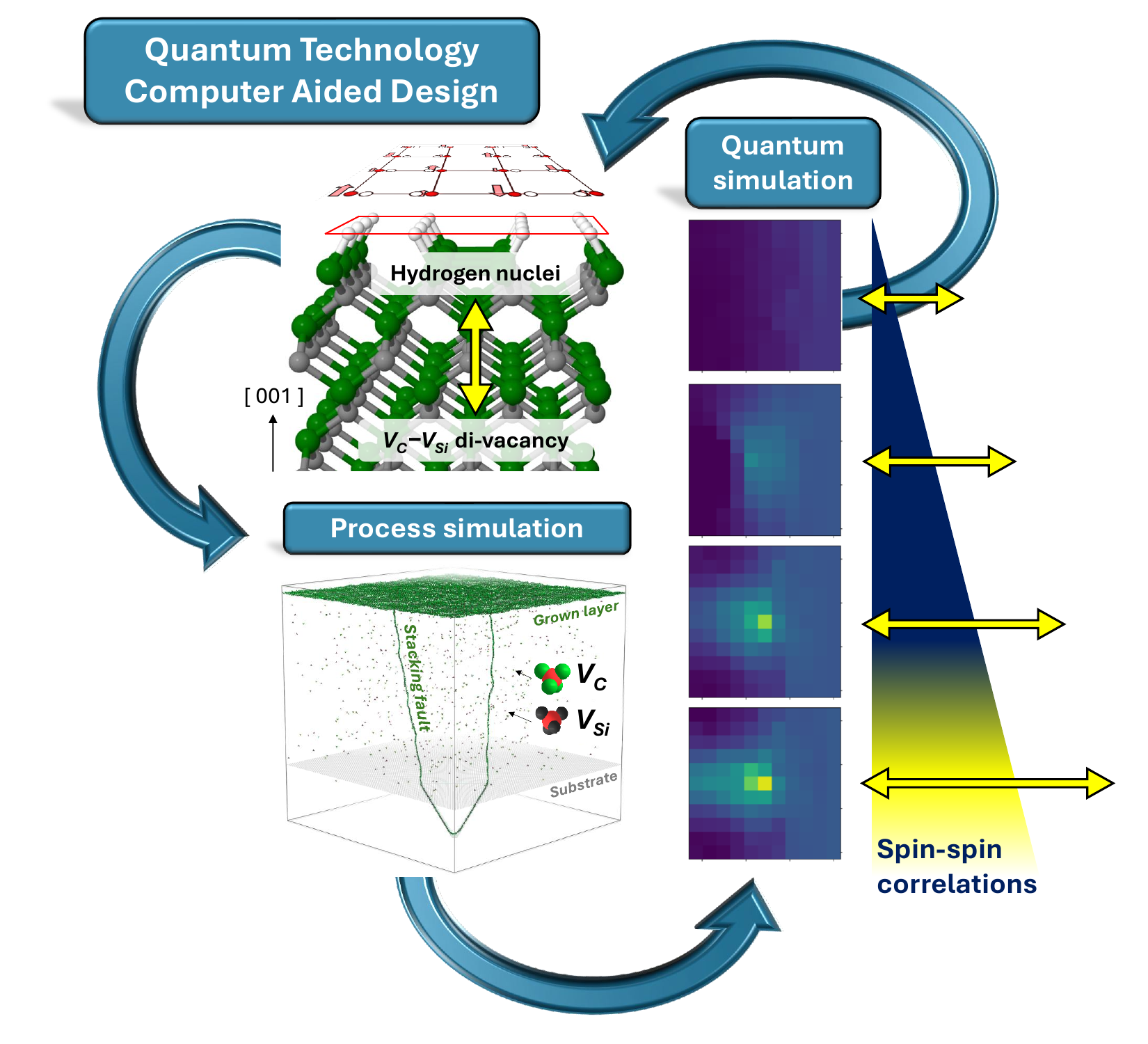}
  \medskip
  \caption*{ToC Image}
\end{figure}

\end{document}